\documentclass[floatfix,aps,physrev,superscriptaddress,twocolumn,10pt]{revtex4-2}

\usepackage{graphicx}
\usepackage{booktabs}
\usepackage{array}
\usepackage{dcolumn}

\usepackage[hidelinks]{hyperref}
\usepackage{orcidlink}

\usepackage{xurl}

\usepackage{xcolor}

\begin{document}


    \title{Skepticism vs. Convenience: Physics Students' Perceptions and Use of Large Language Models Before and After Instruction}

\author{Jake O'Brien \orcidlink{0009-0007-6724-8499}}
\affiliation{Department of Physics, Virginia Tech, Blacksburg, Virginia, USA}
\affiliation{Department of Computer Science, Virginia Tech, Blacksburg, Virginia, USA}

\author{Margaret Ellis \orcidlink{0000-0003-3959-4304}}
\email{Contact author: maellis1@vt.edu}
\affiliation{Department of Computer Science, Virginia Tech, Blacksburg, Virginia, USA}

\author{Alma Robinson \orcidlink{0000-0002-2343-1634}}
\email{Contact author: alrobins@vt.edu}
\affiliation{Department of Physics, Virginia Tech, Blacksburg, Virginia, USA}

\author{John Simonetti \orcidlink{0000-0002-8677-3295}}
\affiliation{Department of Physics, Virginia Tech, Blacksburg, Virginia, USA}

\author{Naren Ramakrishnan \orcidlink{0000-0002-1821-9743}}
\affiliation{Department of Computer Science, Virginia Tech, Alexandria, Virginia, USA}

\date{\today}

\begin{abstract}
    The recent emergence of large language models (LLMs) has produced research focusing on the ability of these tools to solve physics problems, evaluate student work, or otherwise impact the problem-solving process of students. However, studies exploring how physics students perceive LLMs (in terms of capabilities, educational impacts, usage, and role in problem solving) remain limited. This study evaluates the first-year physics students' perceptions toward LLMs and further explores how these perceptions change after practicing problem-solving with and without LLMs and engaging in a reflective lesson on the functioning and educational impacts of LLMs. We find that student opinions toward LLMs vary, with generally favorable perceptions of their capabilities but greater skepticism regarding their value for learning. Despite this skepticism, a majority of students self-report regularly using LLMs to obtain help, commonly reporting deadlines and convenience as motivating factors. Following the lesson, students expressed greater skepticism toward LLMs in several areas, with 88\% of students believing that LLMs can leave them with a false sense of confidence about their understanding, up from 58\% before the lesson.
\end{abstract}


\maketitle

\section{Introduction}
Large Language Models (LLMs) such as ChatGPT from OpenAI have become widely available to physics students. Understanding the ways in which these tools may be used and perceived by students is key to designing effective physics courses in the future. Unlike traditional educational resources, LLMs are capable of providing plausible, personalized responses with little barrier to entry, while simultaneously having no guarantee of correctness or pedagogical intentionality. Effective use of these tools for learning relies on students' ability to critically evaluate the outputs of LLMs and their learning consequences, as well as students' self-regulation skills to avoid over-reliance on LLMs. 

There are a variety of ways that students can use LLMs outside of in-class instruction. Examples include generating additional conceptual explanations, summarizing course materials, or suggesting practice problems. Additionally, students may use LLMs to aid in problem solving or the completion of assignments, such as generating solutions, asking the LLM for guidance/hints, or using the LLM to check their work or thought process.

There are also a variety of reasons why students may choose to use LLMs. Some may use LLMs because they believe these tools will improve their understanding of the material (which may or may not actually be true). Others may turn to LLMs because they feel more comfortable approaching an LLM than a teacher for help, find LLMs more convenient than other resources (teachers/peers/texts/online content), are under pressure to complete assignments quickly, or believe the information generated by an LLM is more clear than the information they can receive from a teacher. 

In order to explore how physics students think about LLMs, and how these perceptions change after instruction and reflection, we investigate the following research questions (RQs): 
\begin{description}
    \item[RQ1] How capable do students think LLMs are at physics problem-solving?
    \item[RQ2] How do students think about the role of LLMs in helping them solve problems?
    \item[RQ3] How do students think about the role of LLMs in learning physics?
    \item[RQ4] What purposes do students use LLMs for, and what purposes do they think they \textit{should} be used for?
    \item[RQ5] How do these perceptions change after instruction on the internals of LLMs and effects they have on learning, and after student reflection on how LLMs affect the learning process?
\end{description}

In this study, we first attempt to establish a baseline regarding how first-year physics students think about the educational impacts of LLMs through a primarily-quantitative, Likert-style survey. We then perform a three-part lesson including a problem-solving activity involving LLMs, introduction to their internal functionality, and a discussion of existing research regarding their impacts on student learning. During each component of the lesson, we promote student reflection through group discussion. Finally, we re-administer the survey after the lesson and observe if/how student perceptions change.

\section{Literature Review}
Several studies have found LLMs capable of solving undergraduate physics problems with high accuracy.
ChatGPT 4 has been demonstrated to show expert-level performance on the Force Concept Inventory (FCI) \cite{West2023}, and was confirmed in another study to correctly and concisely answer all questions in both the FCI and Conceptual Survey of Electricity and Magnetism (CSEM) that do not contain images \cite{Tong2024}. While correct interpretation of graphs and diagrams has been known weakness of LLMs \cite{Kortemeyer2025, Polverini2025}, Kortemeyer observed significant improvement in newer models, with GPT-5 capable of scoring 24 out of 26 \cite{Kortemeyer2026} on the Test of Understanding Graphs in Kinematics \cite{Zavala2017}, as compared to an earlier study which found GPT-4 scores a median of 11 out of 26 \cite{Polverini2024}.

More advanced ``reasoning models'' have also been shown to solve more advanced problems with varying levels of accuracy. A benchmark containing 1,297 physics PhD qualifying exam questions found the best performing ChatGPT model (o3-mini) to achieve an accuracy of 59.9\%, in contrast to human experts who typically achieved scores in the range of 70\% to 80\% \cite{Feng2025}. Another study found that two ChatGPT models (4o and o1-preview) were capable of outperforming human participants on Olympiad-level physics problems \cite{Tschisgale2025}.

It has also been shown that, for introductory physics, LLMs are reasonably accurate at evaluating and providing feedback on student work, even matching the accuracy of human experts when given human-created rubrics and applying self-consistency techniques \cite{Chen2025}.

Other studies investigate the ways students interact with LLMs. Several of these find that students tend to treat LLMs as fact-providers to be forced toward an answer, rather than carefully critiquing them or engaging in deeper discussion as they might with a human peer \cite{Tong2025,Elson2025}. A study allowing students to use ChatGPT as a lab partner found that groups with stronger physics background or stronger understanding of how LLMs function tend to be better positioned to evaluate the trustworthiness of ChatGPT responses, while those with weaker physics background or understanding of LLMs were more likely to simply take ChatGPT answers as being the truth \cite{Kilde2025}.

Studies have also investigated ``moderated'' LLMs, which are designed to overcome some negative repercussions of LLM use by prompting LLMs not to directly reveal answers. Krupp \textit{et al}. found that such moderated LLMs cause more reflective usage than unmoderated models, but still less reflective usage than traditional search engines \cite{Krupp2024}. Another study found that when given access to a moderated LLM in a lab setting, students primarily used the LLM to verify their answers \cite{Kregear2025}.

One concern regarding the availability of LLMs is whether students opt to use them in productive ways, particularly in unsupervised settings. A body of research suggests that many students employ study strategies that feel effective because they help students build familiarity with the material, but students frequently confuse that familiarity with competency \cite{Bjork2013, Dunlosky2013, Mestre2020}. These strategies often require minimal effort and do not engage the student in desirable difficulties \cite{Bjork2011}. Across three studies, Kirk-Johnson \textit{et al}. found that participants who perceived a learning strategy as more effortful rated it as less effective for learning \cite{Kirk-Johnson2019}.

A student's level of engagement affects their learning. For example, studies have shown that the efficacy of learning physics using worked examples is highly dependent on how well the students explain the solution to themselves \cite{Chi1989, RENKL1997}. For example, simply reading the worked solution without a careful self explanation can lead a student to perceive their fluency as competency \cite{Mestre2020}.

Our study and lesson aim to both understand student beliefs regarding the efficacy of LLMs in instruction and foster students' self-regulated learning (SRL) skills, particularly in situations where students may overestimate the effectiveness of their LLM learning strategies. SRL refers to the idea of planning, monitoring, and evaluating one’s learning. Common frameworks describe SRL as consisting of three components: cognition (information/task processing strategies), metacognition (an ability to regulate cognition), and motivation \cite{Muijs2020}. The lesson was designed to support aspects of SRL by encouraging students to reflect on how LLMs influence their problem-solving processes and understanding, and by providing information about how LLMs function that may inform more critical evaluation of their outputs. While the intervention primarily targets metacognitive awareness and reflection, it may also influence students’ selection and awareness of cognitive strategies when engaging with LLMs. 

\section{Method}
\begin{table*}
    \caption{\label{tab:survey-statements}Statements used in the Likert scale portion of the survey, along with their category and the short codes used throughout the paper}
    
    \begin{ruledtabular}
        \begin{tabular}{llp{0.53\linewidth}}
        Category & Code & Statement \\
        \midrule
        Capabilities & \textsc{accuracy-above-students} & ``LLMs are currently better at solving problems than the average physics student.'' \\
        Capabilities & \textsc{accuracy-check-work} & ``LLMs are capable of accurately checking my work for mistakes (e.g by me uploading a picture of my solution).'' \\
        Capabilities & \textsc{accuracy-intro-physics} & ``LLMs can solve introductory physics problems with high accuracy.'' \\
        Comfort & \textsc{comfort-over-classmates} & ``I am more comfortable asking for help from an LLM than from my classmates.'' \\
        Comfort & \textsc{comfort-over-teachers} & ``I am more comfortable asking for help from an LLM than from my instructors or TAs.'' \\
        Learning Impacts & \textsc{helpful-checking-work} & ``LLMs can help me learn by checking my work.'' \\
        Learning Impacts & \textsc{helpful-over-teacher-conceptual} & ``If I need conceptual help, an LLM is more helpful than talking to an instructor/TA.'' \\
        Learning Impacts & \textsc{helpful-over-teacher-problems} & ``If I am stuck on a problem, an LLM is more helpful than talking to an instructor/TA.'' \\
        Usage & \textsc{motivation-convenience} & ``If I did use an LLM, it would likely be caused by it being more convenient than office hours.'' \\
        Usage & \textsc{motivation-deadlines} & ``If I did use an LLM, it would likely be caused by pressure from deadlines.'' \\
        Social Impacts & \textsc{social-reduced-connections} & ``I am worried the availability of LLMs will make it harder to build connections with my classmates by solving problems together.'' \\
        Social Impacts & \textsc{social-reduced-vulnerability} & ``I am worried the availability of LLMs will make students less comfortable being vulnerable and admitting they do not understand a concept.'' \\
        Learning Impacts & \textsc{teacher-false-confidence} & ``Getting help on a problem from an instructor/TA can leave me with a false sense of confidence about my understanding.'' \\
        Learning Impacts & \textsc{usage-dependence} & ``If I over-rely on LLMs in completing my coursework, I may become dependent on them in the future.'' \\
        Learning Impacts & \textsc{usage-false-confidence} & ``Getting help on a problem from an LLM can leave me with a false sense of confidence about my understanding.'' \\
        Learning Impacts & \textsc{usage-net-learning-benefit} & ``Overall, having access to an LLM while doing homework is a net benefit to my learning.'' \\
        Learning Impacts & \textsc{usage-push-careful-thinking} & ``An LLM is more likely to push me to think carefully about problems than an instructor or TA.'' \\
        Learning Impacts & \textsc{usage-reduced-conceptual} & ``If I use an LLM to generate solutions, I am less likely to develop my conceptual understanding.'' \\
        \end{tabular}
    \end{ruledtabular}
\end{table*}
\subsection{Participants \& Course Context}
The study was performed in a required First-Year Experience (FYE) course for physics majors at Virginia Tech. All students were physics majors, and the vast majority were first-year students concurrently enrolled in the second semester of a calculus-based introductory physics course, focusing on oscillations, waves, and electricity and magnetism. The FYE course was co-taught by two instructors, with support from a Graduate Teaching Assistant (GTA) and 2--3 undergraduate Learning Assistants (LAs). The third and fourth authors served as the course instructors, and the first author served as an undergraduate LA during the study.

The LLM policy for the introductory physics course allows their use to assist in conceptual understanding, but not for completing graded homework problems. Beyond this policy, LLMs were not previously discussed in either class.

\subsection{\label{sec:methods-survey}Survey}
Students were given time in class to complete a survey before and after the lesson. To encourage honest responses, students were informed that the survey would be anonymous, and everyone associated with the research study or grading of the course left the room while students had access to the survey.

A total of 65 students were enrolled in the first-year experience course at the time of the lesson, with 57 students being present during both class sessions in which the lesson was taught, and only one student missing both classes. Of those present, 53 students opted to participate in the pre-lesson survey, and 42 opted to participate in the post-lesson survey. We note that the attrition between the pre- and post-surveys creates a risk of sampling bias when measuring changes in aggregate responses. Due to the anonymous nature of the survey, we do not have matched data and cannot limit our analysis to those who completed both surveys. Students were provided no incentive for participation, and the Institutional Review Board determined the research protocol to be exempt from IRB review.


Survey questions primarily consisted of seven-level Likert scale questions based on each of the research questions. We have divided the Likert scale statements into the following categories for our analysis:
\begin{itemize}
    \item Capabilities of LLMs (RQ1)
    \item Learning Impacts of LLMs (RQ2/RQ3)
    \item Social Impacts of LLMs (RQ3)
    \item Usage Reasons (RQ4)
    \item Comfort using LLMs (RQ4)
\end{itemize}

The statements used in the survey are reported in Table~\ref{tab:survey-statements}, along with their categories and a short code referenced in the text of the paper for brevity.

The survey also contained multiple-choice questions designed to compare the frequency with which students use LLMs for different purposes. Students were asked ``For each of the following descriptions, select how often you use LLMs for that purpose.'' and ``For each of the following descriptions, select how often you think LLMs should be used for that purpose to best foster learning.'' with the following descriptions:
\begin{itemize}
    \item Getting conceptual explanations
    \item Getting guidance/hints on problems
    \item Getting solutions to problems
    \item Checking my work or thought process
    \item Summarizing course materials
    \item Generating additional practice problems
\end{itemize}
For each of the above descriptions students were presented with the following options:
\begin{itemize}
    \item Never
    \item Less than once per month
    \item 1-3 times per month
    \item About once per week
    \item Several times per week
    \item Daily
\end{itemize}

Finally, students were also asked to briefly respond to the following qualitative questions:
\begin{itemize}
    \item ``In which ways do you believe LLMs to be most helpful for your own learning, if at all?''
    \item ``In which ways do you believe LLMs to be most detrimental to your own learning, if at all?''
    \item (post-survey only) ``How, if at all, have your opinions on LLMs in learning physics changed after the lesson and in-class activity?''
\end{itemize}

\subsection{Lesson}
After taking the pre-survey, the students participated in a lesson delivered over the course of two 75-minute class sessions. The lesson consisted of three parts:
\begin{enumerate}
    \item A problem-solving activity requiring students to solve physics problems with and without access to LLMs, and a reflection on how LLMs affect their problem-solving process and knowledge retention.
    \item Instruction regarding the internal functioning of LLMs, and some of the techniques that allow language models to solve physics problems.
    \item Instruction highlighting some of the available research on how LLMs impact student learning.
\end{enumerate}

The primary goals of the lesson were to develop the students' ability to critically evaluate the outputs and effects of LLMs, and to develop the self-regulation skills necessary to avoid over-reliance. The capabilities and internal functioning of LLMs were discussed to give context and support these goals, but technical understanding of LLMs was not the primary focus of the lesson. 

\subsubsection{Problem-Solving Activity}
The problem-solving activity was designed to help students reflect on how LLMs impact the problem-solving process, how they affect learning, and how responses generated by LLMs compare to discussion with human instructors and peers.

For the activity, students were presented with a Doppler effect problem (which they had not been taught how to solve) and told to solve it, using an LLM to receive instruction on how to do so. Students were not restricted to using a particular LLM, and instructors observed usage of ChatGPT, Microsoft Copilot, Claude, and Google Gemini. Students were also told that there would be a follow-up problem checking their understanding, and encouraged to take notes for later referencing and ask follow-up questions of the LLMs as needed to develop their understanding.

The initial problem required them to calculate the observed frequency when an observer moves toward a stationary source, and explain conceptually whether the observed frequency would be higher, lower, or equal to the source frequency. The formula for calculating Doppler shifts was not taught in the concurrent introductory physics course, although the conceptual origins of Doppler shifts were briefly discussed.

After solving the problem with LLM assistance, nearly all of the students obtained the correct answers. The students were then asked to ``rate your understanding of how to solve Doppler effect problems from 1 (not confident) to 10 (I’ve got this!)'' on a piece of paper. Collecting the papers, we found a majority of students to rate their confidence at 7 or higher.

Students were then given a second Doppler effect problem, this time with a moving source instead of a moving observer, and asked to solve it using their notes from the previous problem, but without additional LLM access. Once the students had time to attempt the problem, they were allowed to discuss their thinking with peers, instructors, the GTA, and LAs.

Before the discussion period, a majority of students obtained an incorrect answer to the second problem. Furthermore, the instructors, GTA, and LAs observed that many students who obtained incorrect answers appeared confident in their solution. During subsequent discussion, these students often did not recognize their error until the conceptual and mathematical difference between a moving source and a moving observer were explicitly addressed.

After the activity, students were given time to discuss in groups and share their reflections on LLM use with the class. Many students experienced a cognitive mismatch between their high confidence that they could solve the second Doppler effect problem successfully and the feedback indicating their solution was incorrect. This mismatch may have heightened their metacognitive awareness of how the LLM influenced their learning. \cite{butterfield2001,metcalfe2012}.

In the discussion, some themes identified by students were the importance of following up with the LLM to obtain more conceptual details, the fact that LLM responses tended to be helpful and accurate overall, the fact that LLMs may not always highlight certain ideas when solving problems in the same way a human instructor might, and the fact that whether seeking assistance from human instructors or LLMs, it is possible to gain a false sense of confidence if you do not critically engage with the material. Research has shown that an effective instructional strategy is helping students build explanations by asking and answering deep questions \cite{pashler2007}, and students were more likely to engage in this level of questioning with a instructor, GTA, LA, or peers than with the LLM.

\subsubsection{Internal Functioning of LLMs}
Following the activity, the students received instruction regarding the internal functioning of LLMs. Key points included:
\begin{itemize}
    \item The structure of LLMs as probabilistic models that repeatedly predict the most likely words/tokens to continue the text.
    \item The general mathematical structure of a transformer, including the concept of embedding vectors (and how they encode meaning), and the fact that the majority of the prediction process involves using linear algebra to operate on these vectors.
    \item Techniques used to make LLMs more accurate at problem-solving, including chain-of-thought prompting \cite{Wei2022}, self-consistency techniques \cite{Wang2023}, interfacing with external tools (such as python) to improve mathematical accuracy, and appending tokens such as ``wait'' to extend reasoning traces and improve accuracy \cite{Muennighoff2025}.
\end{itemize}

Following instruction on how LLMs work, students were given the opportunity to discuss, in groups, while instructors and TAs circulated. Some points that were commonly cited as being interesting were how LLMs are probabilistic models and how the simplicity of some techniques used to make LLMs more accurate caused opinions to change in regard to their problem-solving skills.

\subsubsection{Educational Impacts of LLMs}
The lesson concluded with discussion of the educational impacts of LLMs. Studies were cited regarding the accuracy of LLMs in solving problems \cite{Feng2025}, the abilities of LLMs to respond to student questions in a lab setting \cite{Kregear_Babayeva_Widenhorn_2025}, the use of LLMs for peer instruction \cite{Tong2025}, and the results of some studies on how LLMs affect student learning in computer science \cite{Vadaparty2024, Kazemitabaar2024, Prather2024} (where more research in this regard has been conducted).

\begin{table*}
\caption{\label{tab:likert-results}Results for the Likert Scale survey questions, including the median scale point (with 1 being strongly disagree, 4 being neutral, and 7 being strongly agree), the percentage of students indicating any level of agreement/disagreement, the Effect Size as calculated by Cliff's Delta, and the $p$-value indicating whether there is a statistically significant difference between the pre- and post-survey results. Note: Effect size is indicated by $^*\left|\delta\right| > 0.15$, $^{**}\left|\delta\right| > 0.33$ and statistical significance is indicated by $^*p<0.05$, $^{**}p<0.01$}
\begin{ruledtabular}
\begin{tabular}{l *{6}{w{c}{1cm}} D{.}{.}{2.4}D{.}{.}{1.5}}
Statement Code
& \multicolumn{2}{c}{Median} 
& \multicolumn{2}{c}{Agreement (\%)} 
& \multicolumn{2}{c}{Disagreement (\%)} 
& \multicolumn{1}{c}{Effect Size} 
& \multicolumn{1}{c}{$p$-value} \\
\cmidrule(lr){2-3} \cmidrule(lr){4-5} \cmidrule(lr){6-7}
& pre & post & pre & post & pre & post & & \\ \midrule
\textsc{usage-push-careful-thinking} & 2 & 2 & 13 & 2 & 73 & 82 & 0.24* & .043* \\
\textsc{helpful-over-teacher-problems} & 3 & 3 & 15 & 10 & 56 & 57 & 0.06 & .605 \\
\textsc{teacher-false-confidence} & 3 & 3 & 27 & 28 & 54 & 57 & -0.00 & .984 \\
\textsc{helpful-over-teacher-conceptual} & 3 & 3 & 29 & 5 & 56 & 72 & 0.17* & .157 \\
\textsc{comfort-over-classmates} & 4 & 4 & 38 & 40 & 40 & 38 & 0.03 & .783 \\
\textsc{comfort-over-teachers} & 4 & 4 & 42 & 45 & 38 & 32 & -0.05 & .701 \\
\textsc{usage-net-learning-benefit} & 4 & 4 & 44 & 50 & 29 & 28 & 0.01 & .929 \\
\textsc{social-reduced-connections} & 4 & 5 & 50 & 72 & 35 & 18 & -0.32* & .007** \\
\textsc{accuracy-above-students} & 5 & 3 & 54 & 32 & 25 & 52 & 0.36** & .002** \\
\textsc{accuracy-check-work} & 5 & 4 & 54 & 48 & 23 & 28 & 0.15 & .225 \\
\textsc{usage-false-confidence} & 5 & 6 & 58 & 88 & 23 & 5 & -0.37** & .002** \\
\textsc{social-reduced-vulnerability} & 5 & 6 & 60 & 75 & 21 & 10 & -0.22* & .062 \\
\textsc{usage-reduced-conceptual} & 5 & 6 & 63 & 85 & 27 & 8 & -0.31* & .008** \\
\textsc{motivation-deadlines} & 5 & 6 & 65 & 88 & 19 & 2 & -0.38** & .002** \\
\textsc{motivation-convenience} & 5 & 5 & 71 & 75 & 15 & 12 & 0.00 & .980 \\
\textsc{accuracy-intro-physics} & 5 & 5 & 79 & 82 & 13 & 8 & -0.14 & .214 \\
\textsc{helpful-checking-work} & 6 & 5 & 77 & 78 & 12 & 18 & 0.26* & .026* \\
\textsc{usage-dependence} & 6 & 7 & 81 & 88 & 10 & 2 & -0.24* & .036* \\
\end{tabular}
\end{ruledtabular}
\end{table*}

\section{Results}
The Likert Scale questions used a seven-point scale, encoded as follows:
\begin{itemize}
    \item 1: Strongly Disagree
    \item 2: Disagree
    \item 3: Somewhat Disagree
    \item 4: Neither Agree Nor Disagree
    \item 5: Somewhat Agree
    \item 6: Agree
    \item 7: Strongly Agree
\end{itemize}

The results of the Likert Scale questions are reported in Table~\ref{tab:likert-results}. Because the scale points are ordinal, we report the median response as a metric to compare which statements had stronger agreement, as suggested by Springuel \textit{et al}. \cite{Springuel2019}. When considering results for each statement, we also report the agreement and disagreement percentage for each statement, which combines all students who responded with scale points 5-7 and 1-3 respectively.

As a measurement of effect size, we report Cliff's Delta \cite{Cliff1993, Meissel2024}, which measures the degree to which the pre-survey and post-survey distributions do not overlap. Unlike comparing overall agreement and disagreement percentages, this approach does not treat differing degrees of agreement or disagreement to be equal. Cliff's Delta is bounded between -1 and 1, with these values indicating no overlap between the two distributions. It is common to interpret these effect sizes as small ($\left|\delta\right| > 0.15$), medium ($\left|\delta\right|> 0.33$), or large ($\left|\delta\right| > 0.47$) \cite{McPadden2017, Chari2019}.

To determine whether any changes observed between the pre- and post-surveys are statistically significant, we include the $p$-value as computed by the Mann-Whitney U Test \cite{Mann1947}, which is appropriate for ordinal data \cite{MacFarland2016}. While the Mann-Whitney U Test is designed for independent samples, other alternatives are not available as we do not have matched data between the surveys. Due to the exploratory nature of the study, which focuses on identifying patterns over confirming any particular hypothesis, we do not correct $p$-values for multiple comparisons. Applying a correction in this context would increase the risk of Type II error \cite{Rothman1990}, obscuring potentially meaningful patterns and risking overcorrection as responses to survey statements are not independent. Instead, we focus on the effect sizes of potentially-significant changes, which should be considered as hypothesis-gathering rather than confirmatory.

\subsection{Pre-Survey Results}
We first present results from the pre-lesson survey, before students have received any instruction regarding LLMs. To provide consistency between the statement results, we only report results for the 52 students who answered every Likert scale question.

\begin{figure}
    \centering
    \includegraphics[width=0.9\linewidth]{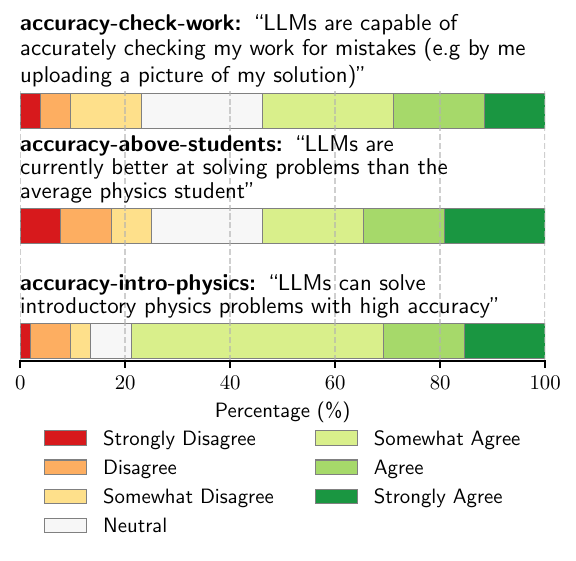}
    \caption{Pre-lesson survey responses to questions regarding the capabilities of LLMs.}
    \label{fig:pre-survey-capabilities}
\end{figure}

\subsubsection{LLM Capabilities}
The results for questions regarding the capabilities of LLMs are shown in Fig.~\ref{fig:pre-survey-capabilities}. The vast majority of students believe LLMs to be capable of solving introductory physics problems with high accuracy (\textsc{accuracy-intro-physics}), with 79\% agreement and 13\% disagreement. There is also a majority agreement that LLMs are capable of accurately checking work for mistakes (\textsc{accuracy-check-work}, 54\% agreement, 23\% disagreement) and that LLMs are currently better at solving problems than the average physics student (\textsc{accuracy-above-students}, 54\%, 25\%).

\begin{figure}
    \centering
    \includegraphics[width=0.9\linewidth]{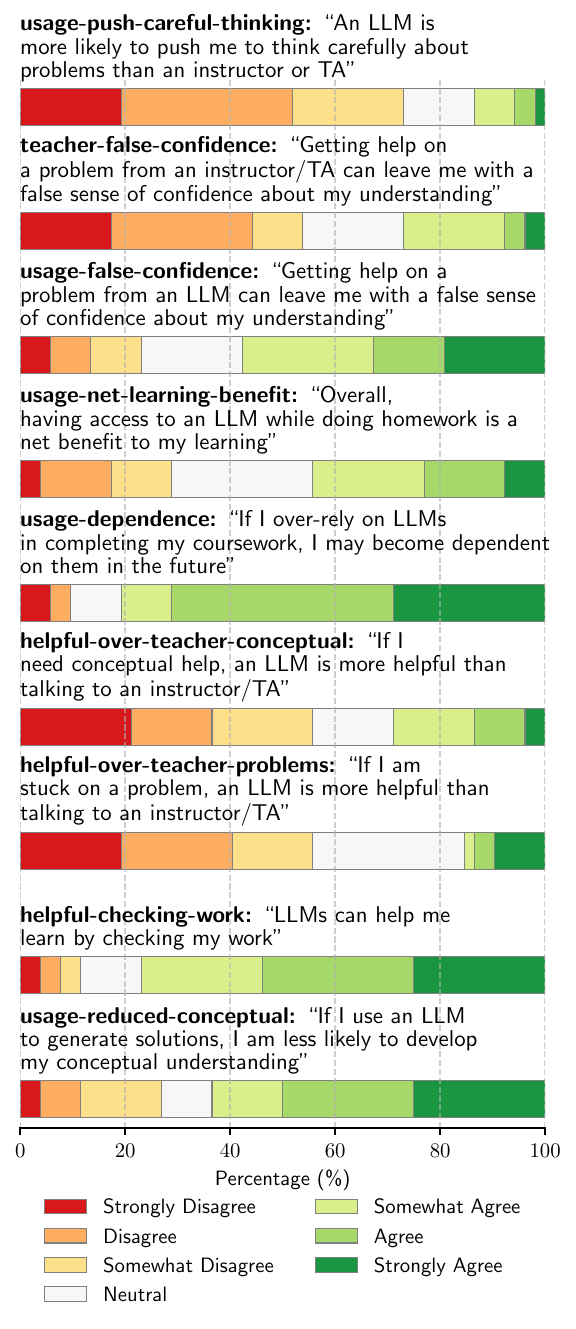}
    \caption{Pre-lesson survey responses to questions regarding the learning impacts of LLMs.}
    \label{fig:pre-survey-learning-impact}
\end{figure}
\subsubsection{Learning Impacts}
The results for questions regarding the learning impacts of LLMs are shown in Fig.~\ref{fig:pre-survey-learning-impact}. The statements regarding the learning impacts of LLMs with majority disagreement were \textsc{usage-push-careful-thinking} (13\% agreement, 73\% disagreement), \textsc{help-over-teacher-conceptual} (29\%, 56\%), and \textsc{help-over-teacher-problems} (15\%, 56\%). We note that all of these results indicate a preference from students to receive explanations from human teachers over LLMs.

The statements related to learning impacts with majority agreement were \textsc{usage-false-confidence} (58\%, 23\%), \textsc{usage-dependence} (81\%, 10\%), \textsc{helpful-checking-work} (77\%, 12\%), and \textsc{usage-reduced-conceptual} (63\%, 27\%). We also found that \textsc{usage-net-learning-benefit} (44\%, 29\%), while having more agreement than disagreement, did not have a majority in either.

Comparing within each student, we find that 67\% of students agree more with the statement ``Getting help on a problem from an LLM can leave me with a false sense of confidence about my understanding'' than with ``Getting help on a problem from an instructor/TA can leave me with a false sense of confidence about my understanding''. 17\% of the students rated the statements equally  and 15\% were in more agreement with the latter statement.

These results indicate that, even without explicit instruction regarding their impacts on learning, our students exhibited some skepticism around using LLMs. While a clear majority of students believe that LLMs can help them learn by checking their work, many express concern that over-use of LLMs may cause dependence, lead to false confidence, and that using LLMs to generate solutions may prevent them from developing conceptual understanding.

We also find that a majority of students indicate that instructors/TAs are more helpful than LLMs for both problem-solving and conceptual help. While this is an interesting result, the nature of the statement means that it may be more indicative of the specific physics courses the students are enrolled in, as opposed to strictly their opinions on the explanatory abilities of LLMs. We discuss the nature of the course further in section~\ref{sec:discussion_pre}.

\begin{figure}
    \centering
    \includegraphics[width=0.9\linewidth]{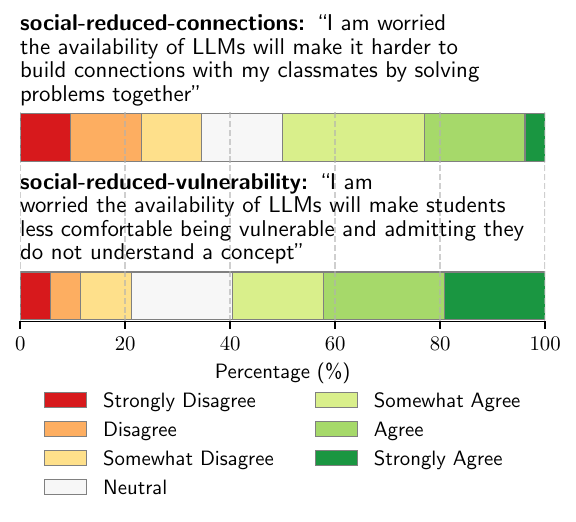}
    \caption{Pre-lesson survey responses to questions regarding the social impacts of LLMs.}
    \label{fig:pre-survey-social}
\end{figure}
\subsubsection{Social Impacts}
For the questions regarding the social impacts of LLMs (shown in Fig.~\ref{fig:pre-survey-social}), we see plurality agreement for both \textsc{social-reduced-connections} (50\% agree, 35\% disagree) and \textsc{social-reduced-vulnerability} (60\%, 21\%).

This indicates that there are many students who both care about the interactions they have with their peers and are worried that LLMs may make these interactions more difficult. This may be a result that is helpful to show to students, as many may not be aware that their peers share similar concerns.

\begin{figure}
    \centering
    \includegraphics[width=0.9\linewidth]{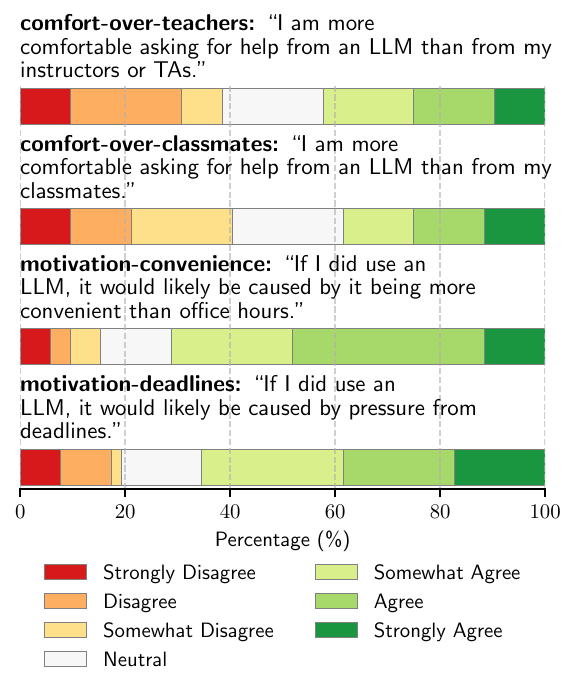}
    \caption{Pre-lesson survey responses to Likert questions regarding the student usage \& comfort with LLMs.}
    \label{fig:pre-survey-usage}
\end{figure}

\subsubsection{Usage \& Comfort}
The Likert scale questions related to usage \& comfort of LLMs are included in Fig.~\ref{fig:pre-survey-usage}. We find a wide spread among students about whether they feel more comfortable getting help from an LLM over talking with humans, with \textsc{comfort-over-teachers} having 42\% agreement and 38\% disagreement, and \textsc{comfort-over-classmates} having 38\% agreement and 40\% disagreement.

Regarding the reasons for using LLMs, we see strong agreement on \textsc{motivation-convenience} (71\% agreement, 15\% disagreement) and \textsc{motivation-deadlines} (65\%, 19\%). The large agreement with these statements, combined with the skepticism toward LLMs observed in previous sections, suggests that there is a significant population of students who may use LLMs even when they do not believe them to be helpful to learning. Furthermore, it appears that convenience and deadline pressure are both significant reasons why this may occur (and are more likely to be motivators than students feeling more comfortable with LLMs than humans).

\begin{figure}
    \centering
    \includegraphics[width=1\linewidth]{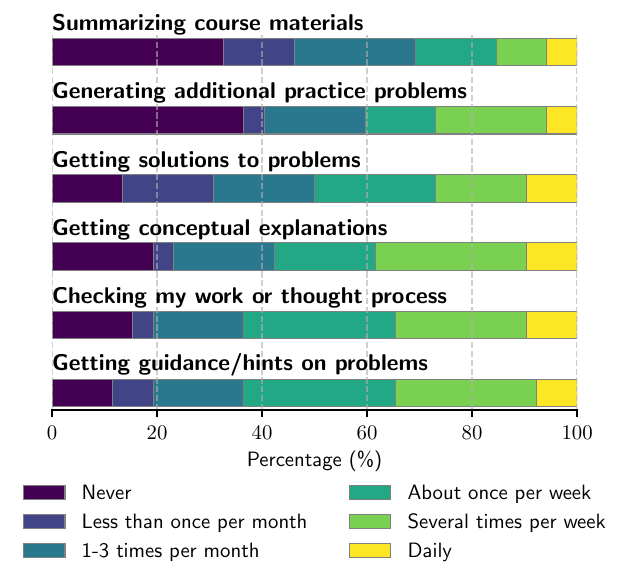}
    \caption{Pre-lesson self-reported LLM usage.}
    \label{fig:pre-survey-reported-usage}
\end{figure}

\begin{table}
\caption{\label{tab:pre-usage-comparison}Comparison between students' pre-survey responses to how often they use LLMs and how often they believe LLMs should be used to best foster learning. Students in the ``More'' column report using LLMs more often for that purpose than what they believe is ideal.}
\begin{ruledtabular}
\begin{tabular}{>{\raggedright\arraybackslash}p{4cm} rrr}
Purpose & More (\%) & Equal (\%) & Less (\%) \\
\midrule
Getting conceptual explanations & 22 & 51 & 27 \\
Getting guidance/hints on problems & 31 & 43 & 25 \\
Getting solutions to problems & 53 & 43 & 4 \\
Checking my work or thought process & 29 & 33 & 37 \\
Summarizing course materials & 18 & 39 & 43 \\
Generating additional practice problems & 8 & 33 & 59 \\
\end{tabular}

\end{ruledtabular}
\end{table}

In addition to the Likert questions, we also had students report how often they use LLMs for a variety of purposes, these results are shown in Fig.~\ref{fig:pre-survey-reported-usage}. Following this, students also reported how often they believe LLMs \textit{should} be used to best foster learning. In Table~\ref{tab:pre-usage-comparison}, we compare the students' responses to measure what portion of students feel they use LLMs more often, less often, or equal to the amount they deem as ideal for learning.

We note that a slight majority of students believe they use LLMs to get solutions to problems more than they should, and a majority of students believe using LLMs to generate additional practice problems more often would be better for their learning.

\begin{figure}
    \centering
    \includegraphics[width=0.95\linewidth]{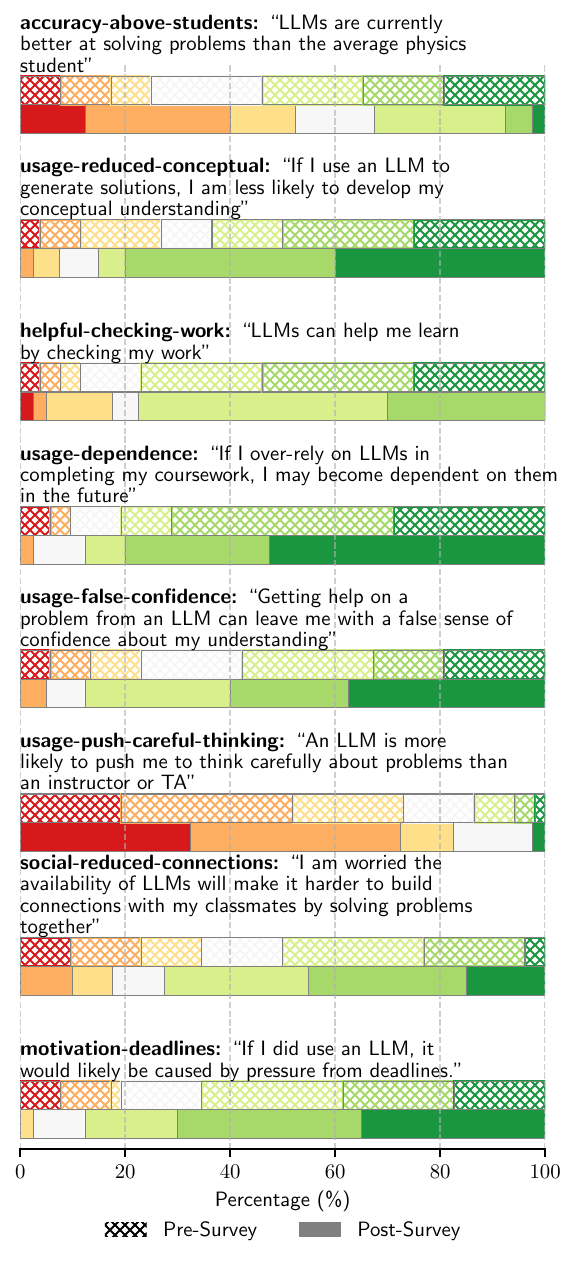}
    \caption{Results for statements which saw a significant ($p<0.05$) shift between pre- and post-survey data. Pre-survey results are included with a hatched background for comparison.}
    \label{fig:pre-post-changes}
\end{figure}
\subsection{Post-Survey Results}
\subsubsection{Likert Scale Questions}
As shown in Table~\ref{tab:likert-results}, eight of the eighteen Likert scale statements saw a statistically significant shift (as defined by a $p$-value of less than 0.05). The post-survey results for these statements are shown in Fig.~\ref{fig:pre-post-changes}, alongside the pre-survey distribution for reference. We note that as discussed in Section~\ref{sec:methods-survey}, the difference in populations between pre- and post-surveys means that shifts may be indicative of differing populations and not necessarily a result of the lesson.

Out of the three statements describing LLM capabilities, we only observe a shift toward disagreement in \textsc{accuracy-above-students}, which had 54\% agreement before the lesson and 32\% agreement after the lesson. The other two statements regarding the capabilities of LLMs (\textsc{accuracy-check-work} and \textsc{accuracy-intro-physics}) remained favorable toward LLMs.

Five of the eight statements that shifted between the surveys related to the learning impacts of LLMs. We found increased agreement in \textsc{usage-reduced-conceptual} (from 63\% to 85\%), \textsc{usage-dependence} (from 81\% to 88\%), and \textsc{usage-false-confidence} (from 58\% to 88\%). There was increased disagreement toward \textsc{usage-push-careful-thinking} (from 73\% to 82\%), with only a single student who agreed with the statement after the lesson. While the change in \textsc{helpful-checking-work} were found to be significant, the changes were marginal, with only a slight increase in disagreement (from 12\% to 18\%).

In regard to social impacts, we found increased agreement toward \textsc{social-reduced-connections} (from 50\% to 72\%), despite the increase in agreement toward \textsc{social-reduced-vulnerability} not being significant. Regarding motivations for using LLMs, we saw \textsc{motivation-deadlines} increase in agreement from 65\% to 88\%. 

\begin{figure}
    \centering
    \includegraphics[width=\linewidth]{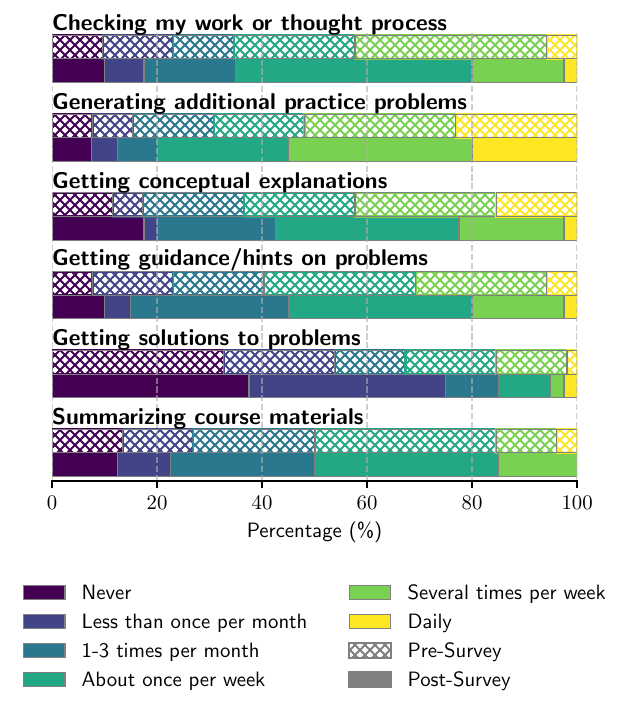}
    \caption{Student responses to the ideal amount of LLM usage to best foster learning. Responses from the pre-survey are indicated with hatched bars, and responses from the post-survey are indicated with filled bars.}
    \label{fig:pre-post-best-usage-comparison}
\end{figure}
\begin{table}
\caption{\label{tab:post-usage-comparison}Comparison between students' post-survey responses to how often they use LLMs and how often they believe LLMs should be used to best foster learning. Students in the ``More'' column report using LLMs more often for that purpose than what they believe is ideal.}
\begin{ruledtabular}
\begin{tabular}{>{\raggedright\arraybackslash}p{4cm} rrr}
Purpose & More (\%) & Equal (\%) & Less (\%) \\
\midrule
Getting conceptual explanations & 38 & 35 & 28 \\
Getting guidance/hints on problems & 35 & 48 & 18 \\
Getting solutions to problems & 75 & 15 & 10 \\
Checking my work or thought process & 38 & 38 & 25 \\
Summarizing course materials & 20 & 30 & 50 \\
Generating additional practice problems & 5 & 25 & 70 \\
\end{tabular}

\end{ruledtabular}
\end{table}

\subsubsection{Usage}
Student responses to the question ``For each of the following descriptions, select how often you think LLMs should be used for that purpose to best foster learning.'' on both the pre- and post-surveys are shown in Fig.~\ref{fig:pre-post-best-usage-comparison}. We found changes between the two surveys generally appear minimal, although we do observe a shift towards students believing that LLMs should be less often used to get solutions to problems.

A similar trend is reflected in the comparison between each student's reported LLM usage, and their beliefs about ideal LLM usage on the post-survey. These results are reported in Table~\ref{tab:post-usage-comparison}. After the lesson, a large portion of students report using LLMs for getting solutions to problems more often than they believe they should (75\% of students, up from 53\% in the pre-survey).

\section{Discussion}

\subsection{\label{sec:discussion_pre}Student Perceptions Prior To Instruction}
From the pre-lesson survey, we observe that student perceptions regarding LLMs are not consistently favorable or unfavorable: context is important.

On the one hand, many students express positive perceptions toward the capabilities of LLMs, with a very large majority of students believing that LLMs can solve introductory physics problems with high accuracy. We also see large agreement that LLMs can be helpful for learning by checking work, and a slight plurality agreement that overall LLMs are a benefit to learning (although the median student is still neutral on this point).

However, we also observe strong agreement on some potential downsides of LLMs, indicating a skepticism about fully embracing the technology. We observed extremely strong agreement toward \textsc{usage-dependence}, with 81\% of students agreeing. We also see strong agreement that LLM use can lead to a false sense of confidence and that the use of LLMs to generate solutions could leave students less likely to develop conceptual understanding.

Despite the skepticism expressed toward the effects of LLMs on learning, we observed 60\% of participants self-reporting using LLMs at least once a week for guidance/hints, 58\% using them at least once a week for conceptual explanations, and 50\% using them at least once a week for problem solutions. We also note that students may feel pressured to downplay the frequency at which they use LLMs, despite the anonymous nature of the survey. This leads to the question of why the majority of students regularly use LLMs, even if they believe that these tools have significant downsides.

When investigating why students turn to LLMs, it appears the reasons are primarily due to convenience and deadlines, which received strong agreement. We saw students were nearly evenly split on whether they were more comfortable talking to an LLM or teacher/classmate, and there was strong majority disagreement toward LLMs being more helpful than instructors/TAs in both conceptual help and problem-solving.

We note that perceptions regarding relative comfort between working with humans and LLMs, and the perceived difference in helpfulness between teachers and LLMs, is likely indicative of student experiences in their physics courses, and not strictly a reflection on their opinions of LLMs. The introductory physics sequence for the vast majority of these students is taught in a SCALE-UP \cite{Beichner2008} classroom with an emphasis on interaction with the instructional team (consisting of instructors, GTAs, and LAs) and peers. We suspect comfort interacting with peers and teachers may be different in traditional large-lecture classes where students are not encouraged to interact with others as often.

\subsection{Student Perceptions After Instruction}

Following instruction regarding LLMs, we observe a large shift toward disagreement in \textsc{accuracy-above-students}. We did not see a significant shift in the other statements assessing LLM capabilities. We find this result to be notable because while the lesson did not attempt to downplay the abilities of LLMs to solve physics problems (and at certain points, the high accuracy of LLMs on introductory physics problems was explicitly noted by instructors), many students' opinions still shifted against LLMs being better problem solvers than physics students. In their post-survey qualitative response, one student wrote ``I don't trust AI as much at all anymore, I didn't know they were probability machines!''. This suggests that discussing the basics of how LLMs function may have been a factor in changing student opinions.

In regard to the effects of LLM usage, we see a significant increase in the agreement to \textsc{usage-false-confidence} (from 58\% to 88\%) and \textsc{usage-reduced-conceptual} (from 63\% to 85\%), demonstrating students are more skeptical of using LLMs for learning following the lesson. It is possible these shifts may have been influenced by experiences students had in the problem-solving activity, where many experienced (or saw peers experience) a sense of false confidence after using an LLM, or noticed the LLMs neglect to discuss conceptual details when not explicitly prompted. This interpretation would suggest there is educational value in having students reflect on LLM use in a classroom environment, even if many would have already tried these tools on their own.

Similar shifts are observed in response to the usage-based questions, where students believed that LLMs should be used less often for getting solutions to problems after the lesson. Other usages (such as for checking work, getting guidance, or summarizing course materials) did not see shifts to the same extent (or at all), indicating that increases in student skepticism towards LLMs for learning was highly dependent on how the tools were used; students did not appear to develop as much concern for strategies involving less cognitive offloading.

We also find the increase in concern about LLMs making it harder to build connections with classmates (\textsc{social-reduced-connections}) interesting, as this was not explicitly addressed as part of the lesson. Perhaps this is caused by students reflecting more on the ways LLMs impact the classroom environment, or it is caused by sampling bias in the fewer number of students that opted to take the post-survey. This is, however, similar to a result that has been previously observed in other disciplines; A qualitative study by Hou \textit{et al}. of computer science students at seven R1 universities finds students report feeling isolated and demotivated as a consequence of the erosion in social support systems caused by LLM access \cite{Hou2025}.

We also observe a significant increase in agreement toward \textsc{motivation-deadlines}, but no significant change in \textsc{motivation-convenience}. One potential cause for this, other than sampling bias, is that students may be less likely to rationalize LLM use as beneficial for learning (as seen in statements such as \textsc{usage-reduced-conceptual}), which pushes them towards justifying their use through deadline pressure. Further research would be beneficial to more closely examine the factors motivating students to use LLMs. 

Overall, we find that almost all shifts in aggregate responses that occurred between the pre- and post-surveys are toward less favorable views on LLMs. This suggests that when given the opportunity to engage with LLMs and reflect on their use, students will not necessarily come to see LLMs as more beneficial for learning, even if they are demonstrated to be effective at answering direct questions (as a number of students observed).

\subsection{Lesson Design \& Observations}

Our interaction with students during the lesson and qualitative results on the post-survey suggest that three components of the lesson (problem-solving activity, discussion of LLM internals, and effects of LLMs on learning) were likely beneficial toward the goal of promoting reflective usage of LLMs.

During the problem-solving activity, we observed a significant portion of the class, after learning how to solve Doppler effect problems from an LLM, express confidence in a wrong answer when asked to solve another problem on their own. We believe this experience to be beneficial, as having firsthand experience of a situation where LLM-based instruction can lead to a false sense of confidence is likely to be more memorable for students than simply being told of this risk.

One concern in the preparation of this activity was that students would view the switch from a moving-observer problem (with LLM access) to a moving-source problem (without LLM access) to be a ``trick'', and not believe their results on the second problem to be a consequence of the LLM's instruction. This concern was mitigated by clearly informing students that they should ensure they prompt the LLM in a way which develops a conceptual understanding, in addition to solving the first problem. We did not observe students expressing perceptions of unfairness in the design of the activity, and after their experience, students seemed willing to consider the idea that LLMs may be less likely to point out conceptual ``pitfalls'' than human instructors when explaining problems.

While we avoided getting too technical for the portion of the lesson describing LLM functionality, we found even the most simple aspects of how LLMs function (namely their probabilistic nature) to be commonly cited by students during discussions as changing their thinking. Our brief discussion of the mathematics behind LLMs, including token embedding examples and a view of the linear algebra involved in a transformer were included with the goal of ``opening the black box'' and reducing the temptation for students to anthropomorphize the models. We wanted students to appreciate that they are fallible machines, notww perfect answer providers. Students also highlighted the discussion of LLM problem-solving techniques (chain-of-thought prompting, self-consistency, interfacing with external tools, and ``wait'' tokens) as affecting their views. On the post-survey, one student wrote ``[their opinions] changed a bit, mostly due to the origin of LLM's `solving' nature'', and another student wrote ``They're a lot less complex than I thought they were.''

For the last portion of the lesson, we discussed some of the ways LLMs affect physics learning. Instead of telling students how to use LLMs, we opted to show a number of studies related to the educational impacts of LLMs as described in the Methods section. This was the most discussion-heavy portion of the lesson, and we spent a significant amount of time giving each table of students a chance to discuss their thoughts and share with the class. One result a number of students highlighted as interesting was the difference in how students performed when collaborating with a human versus an AI, where human-human interactions were found to be the most beneficial \cite{Tong2025}. Other students used this time to discuss how what they observed during the problem-solving activity impacted their overall views on using LLMs to learn, with some expressing how LLMs are best used for asking about specific details or repeated follow-ups.

Overall, we believe that giving students the chance to openly discuss the benefits and drawbacks of LLM use in class is a helpful strategy toward developing more reflective usage of LLMs. Most students will encounter these tools at some point regardless, and giving students the opportunity to compare LLMs and human instructors in an informed manner can allow them become more thoughtful users of LLMs in the future.

\section{Conclusion}

This study provides previously-unexplored insight regarding physics student attitudes towards LLMs. We observe a tension in how physics students engage with large language models: while many recognize the risks of over-reliance, false confidence, and reduced conceptual understanding, they continue to use these tools frequently due to convenience and time pressure.

We find instruction designed to promote reflection on LLM use shifts student perceptions toward greater skepticism, particularly regarding learning impacts, but does not eliminate the underlying motivations for use. The skepticism that develops appears directed towards strategies that involve greater cognitive offloading (such as by asking LLMs to generate solutions) and risks of over-reliance. While this demonstrates some success in addressing the metacognitive component of self-regulated learning, it also shows that additional work is likely needed to address the motivational component.

We suggest that future research should assess interventions focused on improving student motivation for meaningful learning. While we believe that physics students currently express some desire to engage in meaningful learning (as a base concern underlying many concerns around LLM use), this desire is often sidelined when faced with practical pressures such as deadlines and inconvenient alternatives. Potential interventions include reflection on student learning strategies throughout the semester, or having students develop a plan on how they will complete assignments and receive appropriate help before they are assigned \cite{Toro2022,Felker2023}. Such interventions have been implemented in other contexts, and may potentially be beneficial towards encouraging intentional LLM use as well, or reducing unfavorable strategies (such as procrastination) which can push students to use LLMs in ways that are do not effectively support learning.

Finally, future research is also suggested to confirm the patterns we have identified. This study is intended as an exploratory analysis of student perceptions, and future research involving different course contexts (such as with non-physics-majors or those taught in traditional lecture environments over SCALE-UP rooms), larger sample sizes, and matched survey data may help to provide confirmation or nuance to the hypotheses we have identified in our analysis.

\begin{acknowledgments}
This project was supported in part by the Center for Advancing Undergraduate Science Education, (VT-CAUSE), 034212.
\end{acknowledgments}

\section*{Data Availability}
Data supporting this manuscript are publicly available from the Virginia Tech Data Repository \cite{OBrien2026}.

\bibliography{physics_student_llms}

\end{document}